\documentclass{emulateapj}

\begin{document}
\title{Investigating the presence of 500 $\mu$m submillimeter excess emission in local star forming galaxies}
\author{Allison Kirkpatrick\altaffilmark{1}, Daniela Calzetti\altaffilmark{1}, Maud Galametz\altaffilmark{2}, Rob Kennicutt, Jr.\altaffilmark{2}, Daniel Dale\altaffilmark{3}, Gonzalo Aniano\altaffilmark{4}, 
Karin Sandstrom\altaffilmark{5}, Lee Armus\altaffilmark{6}, Alison Crocker\altaffilmark{7}, Joannah Hinz\altaffilmark{8}, Leslie Hunt\altaffilmark{9}, Jin Koda\altaffilmark{10}, Fabian Walter\altaffilmark{5}}
\altaffiltext{1}{Department of Astronomy, University of Massachusetts, Amherst, MA 01002, USA, kirkpatr@astro.umass.edu}
\altaffiltext{2}{Institute of Astronomy, University of Cambridge, Madingley Road, Cambridge, CB3 0HA, UK}
\altaffiltext{3}{Department of Physics \& Astronomy, University of Wyoming, Laramie, WY 82071, USA}
\altaffiltext{4}{Institut d'Astrophysique Spatiale, Universit\'{e} of Paris-Sud, 91405 Orsay, France}
\altaffiltext{5}{Max-Planck Institut f\''{u}r Astronomie, K\''{o}nigstuhl 17, D-69117, Heidelberg, Germany}
\altaffiltext{6}{Spitzer Science Center, California Institute of Technology, MC 314-6, Pasadena, CA 91125, USA}
\altaffiltext{7}{Ritter Astrophysical Observatory, University of Toledo, Toledo, OH 43606, USA}
\altaffiltext{8}{MMT Observatory, University of Arizona, 933 N. Cherry Ave, Tucson, AZ 85721, USA}
\altaffiltext{9}{INAF-Osservatorio Astrofisico di Arcetri, Largo E. Fermi 5, I-50125 Firenze, Italy}
\altaffiltext{10}{Department of Physics and Astronomy, Stony Brook University, Stony Brook, NY 11794, USA}

\begin{abstract}
Submillimeter excess emission has been reported at 500\,$\mu$m in a handful of local galaxies, and previous studies suggest that it could be correlated with metal abundance. We investigate the presence of an excess submillimeter emission at 500\,$\mu$m for a sample of 20
galaxies from the Key Insights on Nearby Galaxies: a Far Infrared Survey with {\it Herschel} (KINGFISH) that span a range of morphologies and metallicities ($12 + \log($O/H$) =7.8-8.7$).
We probe the far-infrared (IR) emission using images from the {\it Spitzer} Space Telescope and {\it Herschel} Space Observatory in the wavelength range $24-500\,\mu$m. We model the
far-IR peak of the dust emission with a two-temperature modified blackbody and measure excess of the 500\,$\mu$m photometry
relative to that predicted by our model. We compare the submillimeter excess, where present, with global galaxy metallicity and, where available, resolved metallicity measurements. We do not find any correlation between the 500\,$\mu$m excess and metallicity. A few individual sources do show excess (10-20\%) at 500\,$\mu$m; conversely, for other sources, the model overpredicts the measured 500\,$\mu$m flux density by as
much as 20\%, creating a 500\,$\mu$m ``deficit".
None of our sources has an excess larger than the calculated 1$\sigma$ uncertainty, leading us to conclude that there is no substantial excess at submillimeter wavelengths at or shorter than 500\,$\mu$m in our sample.
Our results differ from previous studies detecting 500\,$\mu$m excess in KINGFISH galaxies largely due to new, improved photometry used in this study.
\end{abstract}

\section{Introduction}
Dust in the interstellar medium (ISM) is formed by grain growth in both the diffuse ISM and the ejecta of dying stars, such as red giant winds,
planetary nebula, and supernovae \citep{draine2009}.
Once formed, dust absorbs UV and optical light and reemits this radiation in the infrared. 
UV radiation is dominated by photons from young O and B stars, so the dust mass and infrared (IR) radiation can provide important constraints on the current star formation rates and the star formation history
of a galaxy.

Space-based telescopes such as the {\it Spitzer} Space Telescope, the InfraRed Astronomical Satellite, and the Infrared Space Observatory have provided insights on the dust emission from mid-IR wavelengths out to $\sim200\,\mu$m. However,
the majority, by mass, of the dust is cold \citep[$T \lesssim 25$ K,][]{dunne2001} and emits at far-IR and submillimeter wavelengths,
where the emission spectrum is dominated by large grains in thermal equilibrium.
Ground-based observations of these wavelengths are limited to a few atmospheric 
windows, and until recently, space-based observatories lacked coverage of the far-IR beyond $\sim200\,\mu$m. Now, with the advent
of the {\it Herschel} Space Observatory, the peak and long wavelength tail of the dust spectral energy distribution (SED) is being observed at unprecedented angular resolution.

With submillimeter observations, the dust mass and the dust to gas ratios can be more accurately estimated.
Dust formation models show that the dust to gas ratios should be tied to the chemical enrichment of galaxies, as
measured by the metallicity \citep{dwek1998,edmunds2001}. \citet{galametz2011} combine submillimeter data at 450 and 850\,$\mu$m with shorter wavelength IR observations ($\lesssim160\,\mu$m)
for a large sample of local star forming galaxies. They model the spectral energy distributions of each galaxy with and without the submillimeter data and find that for high metallicity galaxies,
not including the submillimeter data can cause the dust mass to be overestimated by factors of 2-10, whereas for low metallicity galaxies ($12 + \log($O/H$) \lesssim 8.0$), the SEDs without the submillimeter data can
underpredict the
true dust mass by as much as a factor of three. \citet{gordon2010} use {\it Herschel} photometry to model the dust emission from the Large Magellanic Cloud (LMC) and find that dust mass derived using just $100\,\mu$m and $160\,\mu$m
photometry underestimate the dust mass derived when using 350\,$\mu$m and 500\,$\mu$m photometry as well by as much as 36\%.

In measuring dust masses by modeling the far-IR and submillimeter SED, a variety of studies have suggested the existence of excess emission at submillimeter wavelengths above what is predicted by fits
to shorter wavelength data. Furthermore, this excess could preferentially affect lower metallicity and dwarf galaxies.
\citet[hereafter D12]{dale2012} report significant excess emission at $500\,\mu$m, above that predicted by fitting the observed SEDs with the \citet{draine2007} models, for a sample of
eight dwarf and irregular galaxies from the Key Insights on Nearby Galaxies: a Far Infrared Survey with {\it Herschel} \citep[KINGFISH,][]{kennicutt2012}. Excess emission has also been reported in the Small
Magellanic Cloud (SMC) and LMC, both of which have lower than solar metallicities \citep[$12 + \log($O/H$)=8.0,8.4$, respectively][]{bot2010,gordon2010}.

On the other hand, excess emission is also seen at solar metallicities or larger. \citet{paradis2012} model emission in the Galaxy and find excess emission ($16-20\%$) at 500\,$\mu$m in peripheral H{\sc ii} regions. Recently, \citet{galametz2013} have detected excess at 870\,$\mu$m on a resolved and global scale for a sample of 11 KINGFISH galaxies. \citet{galametz2011} finds a submillmeter excess at 870\,$\mu$m for 8 galaxies spanning a metallicity range of  $12 + \log($O/H$)=7.8-9.0$.

The presence of excess emission does not appear to be universal, however,
as \citet{draine2007b} found only a marginal difference in deriving the dust masses for a set of SINGS galaxies with and without including submillimeter data from SCUBA $12 + \log($O/H$)\approx7.5-8.7$ . More
recently, \citet{aniano2012} apply the \citet{draine2007} models to the IR SEDs ($3.6-500\mu$m) of the local star forming galaxies NGC 0628 and NGC 6946 and do not detect any significant excess ($>10\%$)
at 500\,$\mu$m.

The excess emission seen at submillimeter wavelengths can be attributed to a cold dust component which is shielded from starlight \citep[T$<$10\,K; e.g.,][]{galametz2009,ohalloran2010}. However, in the Milky Way, excess emission is seen at high latitudes, making it unlikely to be due to shielded cold dust \citep{reach1995}. Studies of other galaxies have argued that a cold dust origin for the excess emission
leads to unphysically high dust to gas ratios \citep{lisenfeld2002, zhu2009, galametz2010}. 

A competing explanation is that the spectral emissivity index, $\beta$, of the dust grains changes to lower values at longer wavelengths,
leading to a flattening of the submillimeter spectrum, thus mimicking a cold dust component \citep{dupac2003, augierre2003,planck2011}. Such a change in emissivity has been suggested in the Milky Way \citep{paradis2009}. \citep{galametz2012} showed that modeling the far-IR /submillimeter emission with a modified blackbody with $\beta=1.5$ can increase the dust masses up to 50\% compared to when $\beta=2$,
possibly explaining the discrepancies between dust masses calculated with far-IR data alone and those calculated with far-IR and submillimeter data.
The emission from very small grains exhibits a frequency dependence with a spectral emissivity index of
$\beta = 1$, and if a galaxy has a relative abundance of very small grains greater than the Milky Way, this can cause a flattening of the emissivity index at longer wavelengths \citep{lisenfeld2002,zhu2009}.
Alternatively, the properties of amorphous solids could explain the change in emissivity \citep{meny2007}. Modeling the emission of amorphous solids accounts for the submillimeter excess in the LMC and SMC \citep{bot2010}, although recent work suggests that much of the excess emission in the LMC is accounted for by fluctuations in the cosmic microwave background \citep{planck2011}. Finally, an increase in the amount of magnetic material has also been proposed as a plausible explanation for the SMC submillimeter excess \citep{draine2012}.

{A change in dust properties can also be linked to metallicity. \citet{draine2007b} showed that $12 + \log($O/H$)=8.1$ is a threshold metallicity for galaxies in the range $12 + \log($O/H$)\approx7.5-8.7$, above which the 
percentage of dust mass contained in PAH molecules drastically increases. If submillimeter excess emission is due to a change in the emissivity properties of the dust population, then there might be a correlation with 
abundances in this metallicity range.

In the present study, we seek to investigate the submillimeter excess at 500\,$\mu$m for a sample of KINGFISH galaxies. We define such excess as being emission above that predicted by a two-temperature modified blackbody model where the emissivity is not constrained, following the method outlined in \citet{galametz2012}. We have selected a sample that spans a range of metallicities, including many dwarf galaxies, in order to probe whether
such excess correlates with metallicity, as has been found in some earlier studies \citep{galametz2009,galametz2011}. The paper is laid out as follows: in Section 2, we describe our sample selection and modeling of the dust emission; in Section 3, we report
on the excess emission and how our findings relate to the results of D12; and in Section 4, we present our conclusions.

\section{Data Analysis}
\subsection{The KINGFISH Sample}
The KINGFISH sample \citep{kennicutt2012} was selected to include a wide range of luminosities, morphologies, and metallicities in local galaxies. The sample overlaps with 57 of the  galaxies observed as part of the Spitzer Infrared Nearby 
Galaxies Survey \citep[SINGS,][]{kennicutt2003b}, as well as incorporating NGC 2146, NGC 3077, M 101 (NGC 5457), and IC 342 for a total of 61 galaxies. The luminosity range spans four orders of magnitude, but all galaxies have $L < 10^{11}\,L_\odot$.
While some of the galaxies display a nucleus with LINER or Seyfert properties, no galaxy's global SED is
dominated by an AGN.

We select a sample (7) of disk galaxies possessing a large angular size ($\sim$30\,arcmin$^2$) and a known metallicity gradient with which we perform a resolved study of the excess emission and search for any correspondence with 
metallicity. We refer to this sample as the ``extended'' galaxies.
Since excess emission seems to be found in dwarf/irregular galaxies, we include the dwarf/irregular galaxies (9) in KINGFISH for which we have a signal-to-noise ratio (SNR) $>3\sigma$ in all of the SPIRE 
bandpasses.
The dwarf/irregulars are more compact objects than our normal disk galaxies, and in general, do not have a known metallicity gradients, so we measure the excess emission globally for these objects. The 
resolved nature
of the normal disk galaxies might bias our comparison with the dwarf/irregular galaxies, so we complement our sample (4) with normal disk galaxies that have small angular size which allows for global measurements 
of excess emission. These more compact, normal galaxies were selected to span a range of metallicity and have high SNRs at the SPIRE bandwidths. Our complete sample consists of 20 galaxies and is listed
 in Table \ref{tbl:sample}. We also list in Table \ref{tbl:sample} dwarf/irregular galaxies that were rejected due to low SNRs.

\begin{deluxetable*}{llcclcccc}
\tablecolumns{9}
\tablecaption{Basic properties of our sample \label{tbl:sample}}
\tablehead{\multicolumn{1}{c}{ID} & \colhead{Morph.\,\tablenotemark{a}} & \colhead{Type\tablenotemark{b}} & \colhead{Dist. \,\tablenotemark{c}} & \multicolumn{1}{l}{$12+\log\left(\frac{\rm O}{\rm H}\right)$\,\tablenotemark{d}} &
\colhead{Axis Ratio\,\tablenotemark{e}} & \colhead{Reg. Size\tablenotemark{f}} &\colhead{Aper. \tablenotemark{g}} & \colhead{\# of Aper.\tablenotemark{h}} \\
\colhead{ } & \colhead{ } & \colhead{ } & \colhead{(Mpc)} & \colhead{} & \colhead{} & \colhead{(kpc$^2$)} & \colhead{} & \colhead{}}
\startdata
NGC 0337   & SBd & C & 19.3  & 8.18 & 1.61 & 107 & 1' 38'' & 1  \\
NGC 0628   & SAc & E & 7.20  & 8.35 [$8.43-0.27\rho$] & 1.11 &  1.85 & 42'' & 48  \\
NGC 0925   & SABd & E &  9.12  & 8.25  [$8.32-0.21\rho$] & 1.78 & 4.95 & 42'' & 42 \\
NGC 1377 & S0 & C & 24.6  & 8.29 & 2.00 & 117 & 1' 10'' & 1  \\
NGC 1482 & SA0 & C & 22.6  & 8.11 & 1.79 & 84.8 & 1' 10'' & 1  \\

NGC 2915 & I0 & D & 3.78  & 7.94 & 1.90 & 2.67 & 1' 10'' & 1  \\
NGC 3077 & I0pec & D & 3.83 & 8.64 & 1.20 & 1.60 & 1' 10'' & 1  \\

NGC 3190 & SAap & D & 19.3 & 8.49 & 2.93 & 39.3 & 1' 52'' & 1  \\
NGC 3198 & SBc & E & 14.1 & 8.34 [$8.49-0.50\rho$] & 2.58  & 18.9 & 42'' & 19  \\
IC 2574      & SABm & D & 3.79 & 7.85 & 2.44 & 1.25 & 42'' & 7  \\
NGC 3773 & SA0 & C & 12.4  & 8.43 & 1.20 & 7.58 & 48'' & 1  \\

NGC 4236 & SBdm & D & 4.45 & 8.17  & 3.04 & 2.34 & 42'' & 25  \\
NGC 4321 & SABbc & E & 14.3  & 8.50 [$8.61-0.38\rho$] & 1.17 & 7.85 & 42'' & 55  \\
NGC 4559 & SABcd & E & 6.98  & 8.29 [$8.32-0.10\rho$] & 2.43 & 3.90 & 42'' & 32  \\ 
NGC 4625 & SABmp & D & 9.30 & 8.35 & 1.06 & 9.03 & 1' 10'' & 1 \\
NGC 5055 & SAbc & E & 7.94  & 8.40  [$8.59-0.63\rho$] & 1.75 & 3.63 & 42'' & 73  \\
NGC 5398 & SBdm & D & 7.66  & 8.35 &1.65 & 17.3 & 1' 38'' & 1  \\
NGC 5408 & IBm & D & 4.80 & 7.81 & 2.00 &  2.84 & 56'' & 1 \\
M 101         & SABcd & E & 6.70  & 8.68 [$8.76-0.90\rho$] & 1.07 & 1.54 & 42'' & 180 \\

NGC 5713 & SABbcp & D & 21.4 & 8.24 & 1.12 &119 & 56'' & 1 \\
Ho I             & IABm & D & 3.90 & 7.61 & 1.20 & 3.33 & 1' 38'' & 1  \\
Ho II	           & Im  & D & 3.05   & 7.72 & 1.25 & 0.380 & 42'' & 1  \\
DDO 053   &  Im & D & 3.61  & 7.60 & 1.15 & 0.870 & 56'' & 1  \\
DDO 154   & IBm & D & 4.30 & 7.54 & 1.36 & 1.49 & 56'' & 1 \\

DDO 165   & Im & D & 4.57 & 7.63 & 1.84 & 2.28 & 56'' & 1  \\
M81 DwB  & Im & D & 3.60 & 7.84 & 1.50 & 0.640 & 42'' & 1  \\

\enddata
\tablenotetext{}{We do not calculate an excess for Ho I, Ho II, DDO 053, DDO 154, DDO 165, and M81 DwB since we are not able to measure flux densities at the 3$\sigma$ level covering the full wavelength
range $24-500\,\mu$m. The metallicity range for which we can derive excess measurements is then $12 + \log($O/H$)=7.8-8.7$.}
\tablenotetext{a}{Morphology as listed in the NASA Extragalactic Database (NED).}
\tablenotetext{b}{We indicate how we have classified the galaxies in this study. E = extended, C = compact, and D = dwarf/irregular/Magellanic.}
\tablenotetext{c}{Distances are taken from \citet{kennicutt2012} and references therein.}
\tablenotetext{d}{Characteristic metallicity for most galaxies is taken from \citet{moustakas2010} where the metallicity is calculated using the method of \citet{pilyugin2005}. For the extended galaxies, we list the metallicity gradients in the brackets, where $\rho$ is the radius in arcminutes.
The metal abundance of NGC 1377 is calculated using the luminosity-metallicity relationship \citep{kennicutt2012}, the abundance for NGC 3077 is empirically calculated in \citet{storchi1994}, and the metallcity for M 101 is derived from observations in \citet{kennicutt2003} and \citet{bresolin2004}.}
\tablenotetext{e}{The ratio of the major axis to minor axis, which we use to correct for inclination effects when calculating the metallicity.}
\tablenotetext{f}{The physical region size of the photometry aperture.}
\tablenotetext{g}{The diameter (in arcminutes and arcseconds) of the circular apertures used for photometry.}
\tablenotetext{h}{The number of circular apertures used to calculate the photometry.}
\end{deluxetable*}

\subsection{Data Reduction}
\label{sec:reduction}
Data observations and reduction are discussed in detail in \citet{engelbracht2010}, \citet{sandstrom2010} and \citet{kennicutt2012}. In the present study, 
we use images from the PACS and SPIRE instruments on the {\it Herschel} Space Observatory spanning a wavelength range of $70-500\,\mu$m and MIPS 24\,$\mu$m images from 
the {\it Spitzer} Space Telescope \citep{kennicutt2003b}. We use PACS 70 and 160\,$\mu$m images instead of MIPS due to the better resolution provided by PACS. All galaxy maps are at least 1.5 times the diameter of 
the optical disk, allowing us to probe the cold dust emission beyond the optical disk.

The raw PACS and SPIRE images were processed from level 0 to 1 with {\it Herschel} Interactive Processing Environment (HIPE) v.\,8. The PACS and SPIRE maps
were then created using the IDL package {\it Scanamorphos} v.\,17 \citep{roussel2011}. 
{\it Scanamorphos} is
preferred to HIPE for its ability to better preserve low level flux density, reduce striping in area of high background, and correct brightness drifts caused by low level noise.
The maps are converted from units of Jy beam$^{-1}$ to MJy sr$^{-1}$ by accounting for beam sizes
of 469.1, 827.2, and 1779.6 arcsec$^2$ for the 250, 350, and 500$\,\mu$m maps, respectively. The design and performance of the SPIRE instrument is discussed in detail in \citet{griffin2010}. In the SPIRE images used in this study, the full width at half maximum (FWHM) of the beam profile has been
modeled with a wavelength dependence, so that FWHM $\propto\lambda^{\gamma}$, where $\gamma=0.85$. As part of the calibration process, a ``color-correction'' has been applied to each beam 
so that the
measured specific intensity $I_\nu$ will be equal to the actual specific intensity $I_0$ when $\nu=\nu_0$. The correction factors applied to these images assume a power-law spectrum with $\alpha=-1.9$.

We compare our results to those of D12, so it is crucial to explicitly state that we are using different versions of the PACS and SPIRE images. In D12, the raw PACS and 
SPIRE images were processed from level 0 to 1 with HIPE v.\,5.
The SPIRE data were then mosaicked using the mapper in HIPE. The images were converted to MJy sr$^{-1}$ using beam sizes of 423, 751, and 1587 arcsec$^2$ at 250, 350, and 500$\,\mu$m, respectively, and the FWHM was not modeled with a wavelength dependence; furthermore, the color-correction was not included in the beam size, but was applied a posteriori.The 
PACS data were mapped using {\it Scanamorphos} v.\,12.5.

The main difference between the version of KINGFISH images that we use and the previous version used by D12
is that the SPIRE 250 and 350\,$\mu$m images contain slightly less flux density ($\sim$2\%) while the 500\,$\mu$m image is less luminous by $\sim$10\%. The point response function area is also
slightly larger ($\sim$3\%) than for the v.\, 2 images. The more recent HIPE v.\,11 pipeline has revised the SPIRE beam sizes in the direction of yielding lower fluxes for extended sources by $\sim$6\%, 6\%, and 8\%
at 250, 350, and 500\,$\mu$m, respectively. This is in the direction of reinforcing the results that are presented in this paper.

\subsection{Background subtraction and Convolution of Images}
\label{sec:background}
The background subtraction is described in detail in \citet{aniano2012}, and we will briefly summarize here. ``Non-background" regions are determined in all cameras (IRAC, MIPS, PACS, and SPIRE)
by those pixels which have a SNR $>$ 2. Masks of the non-background regions are made, and then the masked images from all cameras are combined in order to unambiguously determine which regions are truly background regions.
The signals in the background regions are then averaged, smoothed, and subtracted according to an algorithm outlined in \citet{aniano2012}.

Since we compare our results with those of D12, it is worthwhile to note that a different procedure for background subtraction is followed in that study. The authors use a set of sky apertures to measure the local sky around each galaxy while avoiding any contamination from galaxy emission. The
mean sky level per pixel is computed from these sky apertures, scaled to the number of pixels in the galaxy photometry aperture, and the result is subtracted from the overall galaxy photometry aperture counts.

In order to consistently measure photometry for each galaxy, we used images that had been convolved to the resolution of the SPIRE 500\,$\mu$m bandpass. The convolution was done with publicly available kernels from 
\citet{aniano2011} which transform
the point source functions (PSFs) of individual images to the PSF of the SPIRE instrument at 500\,$\mu$m (FWHM of 38'').
The convolution kernels and methodology are described in detail in \citet{aniano2011, aniano2012}. After 
the convolution to a common PSF, the images for each galaxy are resampled to a standard grid, where each pixel is $\sim$ 14''.  


\subsection{Photometry and SED fitting}
\label{sec:fit}
For the extended galaxies, we measure photometry in apertures with diameters of 42'', chosen to be slightly larger than the beam size of the SPIRE 500\,$\mu$m bandpass, from the center of the galaxy to the 
outskirts.
We reject regions of galaxies without at least 3$\sigma$ flux density measurements at every bandpass from $24 - 500\,\mu$m.
For the galaxies which are compact at the 500\,$\mu$m resolution, including the dwarf/irregulars,
we measure the photometry globally in apertures with diameters ranging from 42'' to 1' 52'', which is the angular area covered by the galaxies with a signal to noise ratio $> 3\sigma$. The diameter and the physical 
scales of each photometric aperture are listed in Table \ref{tbl:sample}.

\begin{figure*}
\plotone{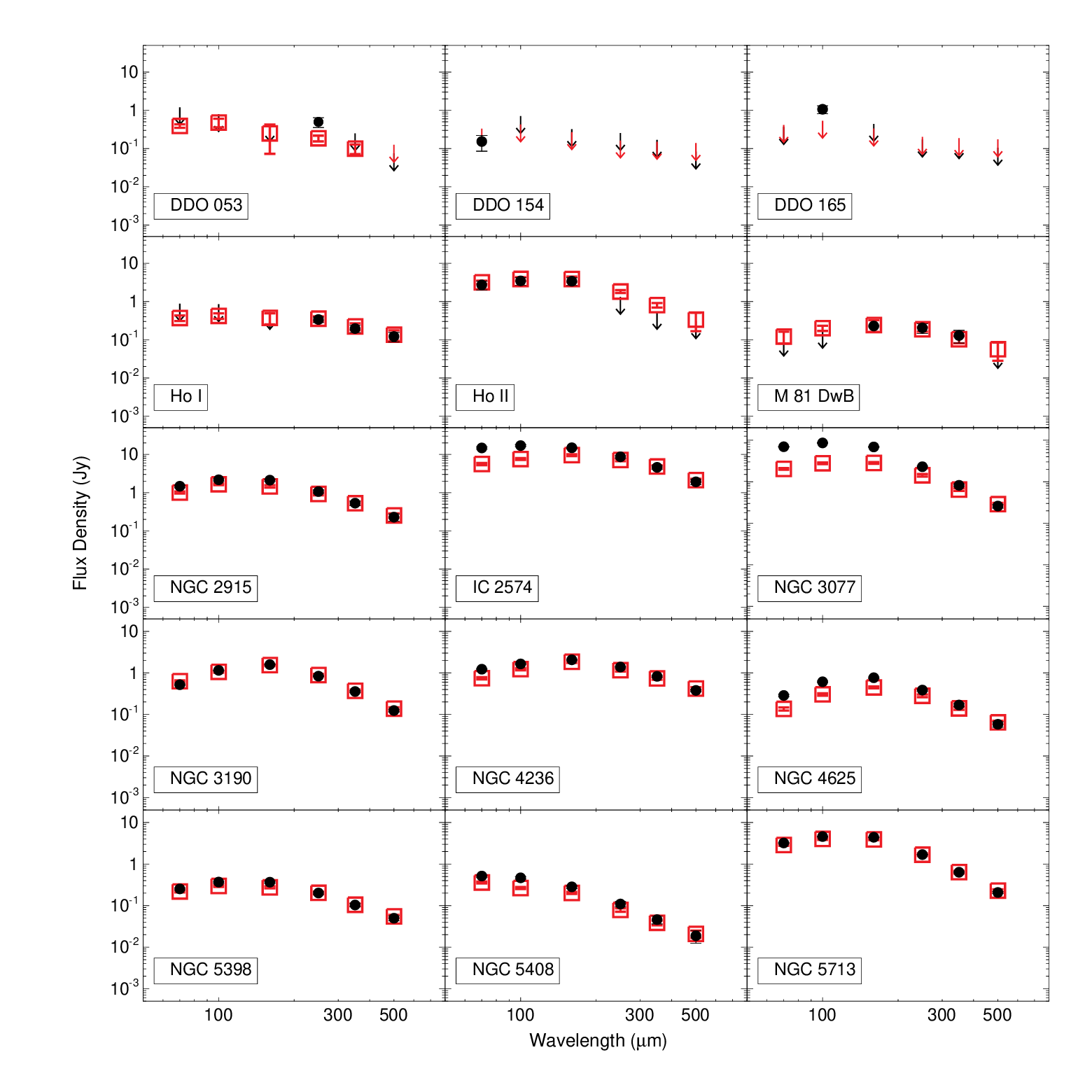}
\caption{Our photometry for the dwarfs/irregular galaxies in the KINGFISH sample is plotted as the black circles. The photometry listed in D12 is over plotted as the red squares. The two sets of photometry were 
extracted from slightly different calibrations of the PACS and SPIRE images, and our photometry was taken after the images had been convolved to the 500\,$\mu$m resolution. When we fit the D12 photometry with our 
model, we are able to reproduce the excesses reported in that work.\label{fig:dale}}
\end{figure*}

The dwarf/irregular galaxies DDO 053, DDO 154, and DDO 165 have a 3$\sigma$ flux density measurement in only one of the far-IR bandpasses; Ho I, Ho II, and M 81 DwB have measurements in three of the six bandpasses. We do not model the far-IR emission for these galaxies, but we do plot the photometry and 3$\sigma$ upper limits in Figure \ref{fig:dale}.

The physically-based DL07 models have been used to derive dust properties at global and local scales \citep[e.g.,][]{draine2007b, aniano2012}. The DL07 models use the assumption of a diffuse
ISM component  to describe the shape of the interstellar radiation field, and use distribution function to represent the distribution of starlight intensities, which are then scaled to the SED being modeled. In addition, the submm slope of the DL07 models tends to resemble that of a modified blackbody with an emissivity
of $\beta=2$, which could prevent us from testing for flattening of the submm slope or variations in the emissivity index.
We choose to model the far-IR SEDs with a simple modified blackbody equation to test various assumptions on the emissivity index and will compare our results to the D12 study to investigate how
the choice of model affects the observed excess emission.

We model the far-IR SEDs of each region within each galaxy using a two temperature modified blackbody of the form
\begin{equation}
\label{eq:bb}
F_\nu= a_w \times B_\nu(T_w) \times \nu^{\beta_w} + a_c \times B_\nu(T_c) \times \nu^{\beta_c}
\end{equation}
where $B_\nu$ is the Planck function.
The temperatures of the warm and cold dust components are $T_w$ and $T_c$, the scalings for each component are $a_w$ and $a_c$, and $\beta_w$ and $\beta_c$ are the emissivity indexes. Only the
scalings, $T_c$, and $\beta_c$ are allowed to be free parameters, due to the limited number of data points being fit.
The warm dust component is necessary to account for some of the emission at 70\,$\mu$m and 100\,$\mu$m so as not to bias the cold dust temperature or emissivity.  When only one temperature is used, the peak of 
the modified blackbody is biased towards shorter wavelengths, leading to warmer
temperatures, which is then compensated at longer wavelengths by a $\beta_c$ shallower than otherwise measured.
We include the 24\,$\mu$m data point in the fit to better constrain the warm dust modified blackbody, but the fitted warm dust parameters should not be interpreted in a physical manner, since this portion of the SED
also likely contains contribution from stochastically heated dust. We list the fraction of 100\,$\mu$m flux density due to the warm dust component in Table \ref{tbl:fit}.

The effective dust emissivity we derive comprises the intrinsic spectral emissivity properties of dust and the variation of the dust temperature within a resolution element that could lead to a 
shallower $\beta_c$ than that obtained in isothermal cases \citep[e.g.,][]{shetty2009}. We nevertheless allow both the cold temperature and cold emissivity, $\beta_c$, to vary simultaneously, though the two
parameters are degenerate. We refer to \citet{galametz2012} and \citet{kirkpatrick2013}, and references therein, for a further discussion of degeneracies on the individual temperatures and emissivities.
 
 We fit the photometry from $24 - 350\,\mu$m. The 500\,$\mu$m data are not included to allow comparison with our model predictions at the same wavelength. We use a Monte Carlo technique to determine the parameters and errors. We randomly sample each data point within its errors 1000 times and determine the parameters of the two temperature modified blackbody via $\chi^2$ minimization.
 The final parameters and associated errors are the medians and standard deviations of our Monte Carlo simulation. An example of the SED fitting is shown in Figure \ref{fig:SED}.
 
 When fitting, we allow the cold dust temperature, $T_c$, to vary between 0 and 50\,K and $\beta_c$ to vary between 0 and 5.
 We experimented with allowing $T_w$ to vary, but the derived temperature was approximately constant 
in the range $55-60$\,K, and simultaneously varying $T_w$ did not change the derived cold dust temperature or emissivity values. Since the warm dust component peaks in an area of the SED that is sparsely sampled,
we opt to hold both $T_w$ and $\beta_w$ fixed. We hold the emissivity of the warm dust component fixed to a value of two and $T_w$ fixed to 60\,K. Fixing $\beta_w=2$ is a good approximation of the opacity of graphite/silicate dust models \citep{li2001}. \citet{galametz2012} find that changing $\beta_w$ to 1.5 decreases the cold dust temperatures by less than 1.6\%, and \citet{tabatabaei2011} test the two temperature modified blackbody approach on M33, holding $\beta_w$ fixed to 1.0, 1.5, and 2.0, and conclude that $\beta_w=2$ most accurately reproduces the observed flux densities.  We test a fixed $\beta_c$ in Section 3.3. As
 $\beta_w$ is always fixed to a value of two in all of our subsequent analysis, in the remainder of this paper, we refer to $\beta_c$ simply as $\beta$.

\begin{figure}
\centering
\plotone{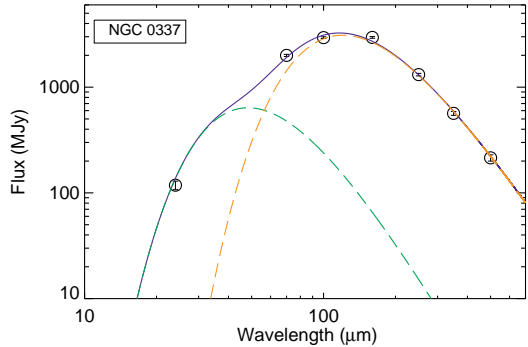}
\caption{The SED of NGC 337 is fitted with a two-temperature modified blackbody. The warm temperature component is shown in green, the cold temperature component in orange, and the composite in blue.
The excess emission is calculated relative to the composite fit at 500\,$\mu$m. The 24\,$\mu$m photometric data is from {\it Spitzer} MIPS, while the other far-IR/submillimeter photometric data are from {\it Herschel}
PACS and SPIRE.\label{fig:SED}}
\end{figure}

\section{Results}
\subsection{PACS-SPIRE colors}
The color $S_{100}/S_{160}$, where $S_{100}$ is the flux density in the 100\,$\mu$m bandpass, is commonly used as a proxy for temperature, since it usually spans the peak of the blackbody emission, while $S_{350}/S_{500}$ depends sensitively on the slope of the Rayleigh-Jeans tail, and so is a good proxy for $\beta$ (with the same caveats given earlier about the mixing of dust with different temperatures along the line of sight). We plot these two colors in the left panel of Figure 
\ref{fig:color} for each of the regions in our extended galaxies. We calculate the positions of theoretical modified blackbodies and overplot as a grid. When calculating the theoretical tracks, we use a two temperature
modified blackbody (Equation \ref{eq:bb}); we set $T_w= 60\,$K, $\beta_w = 2$, and we scale the warm dust component to peak at a flux density 10\% of the cold dust component peak flux density, which is the average scaling
we see in our fitted SEDs at the SPIRE wavelengths. We set $\beta=1,1.5,$ and 2, and we set $T_c$ to discrete temperature values between 15 and 27\,K. Lines of constant $\beta$ and $T_c$ are marked on the figure.

The colors of our galaxies tend to cluster around the $\beta=2, T_c=20\,$K lines. At higher temperatures, galaxy colors start to sparsely occupy the same part of color space as the $\beta=1$ model. No galaxies exhibit colors indicating a low temperature ($T_c<21\,$K) and a shallow emissivity, which could hint at physical link between
the two parameters \citep[see][for a discussion of the physical nature of the $T_c-\beta$ relationship]{yang2007, ysard2012, kirkpatrick2013}. Although our galaxy colors tend to lie near the $\beta=2$ model line,
we observe a large scatter. This illustrates a significant difference of physical conditions from one object to another but also within our objects. Letting both parameters vary in our multiple blackbody approach can be a way to probe these variations, in spite of the degeneracies.

In the right panel of Figure \ref{fig:color}, we plot the same colors, but this time we use the predicted 500\,$\mu$m flux density, which has been calculated by fitting Equation \ref{eq:bb} to the $24-350\,\mu$m flux densities.
Substituting the predicted 500\,$\mu$m flux density has two interesting effects. First, it increases the amount of scatter visible in the colors. For example, in both NGC 5055 and NGC 4321, which display the largest increase
in scatter, the mean $S_{350}/S_{500}$ observed ratio and predicted ratio is approximately the same ($\sim2.7$). For NGC 5055, the standard deviation increases from 0.16 to 0.24, and for NGC 4321 it increases from
0.11 to 0.23. This increase is indicative of the uncertainties
inherent in fitting modified blackbody models without enough data to adequately constrain the slope of the Rayleigh-Jeans tail. The second effect we see is that now the galaxy colors, particularly for M 101, occupy a region of 
the color space to the right of the $\beta=2$ model line. The predicted $S_{350}/S_{500}$ ratio is larger than the observed $S_{350}/S_{500}$ ratio.
Again, this illustrates the importance of using data above 350\,$\mu$m to constrain the submillimeter SED. Without including the 500\,$\mu$m data point in the
modeling of the far-IR/submillimeter SED, the predicted slope will be steeper than the true value; in other words, the model underpredicts the measured 500\,$\mu$m flux density. 

\begin{figure*}
\plotone{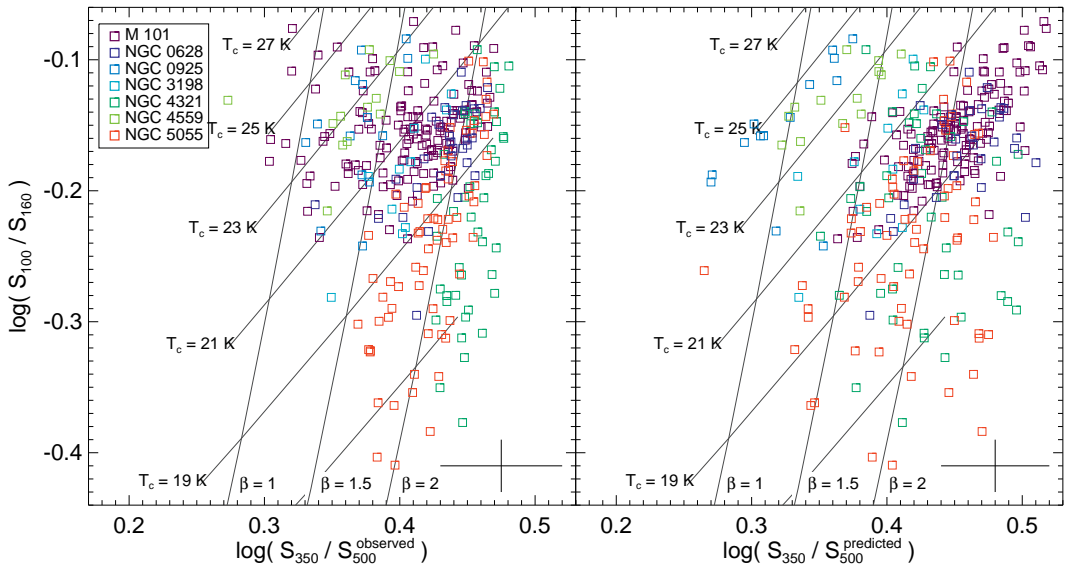}
\caption{{\it Left--} We plot the far-IR colors of each region in our extended galaxies. We calculate the theoretical colors of a two-temperature modified blackbody for different temperatures and different values 
any regions with low temperatures and low emissivity values, hinting at a physical link between the two parameters. The typical errors are shown in the lower righthand corner.
{\it Right--} We replace the observed 500\,$\mu$m flux densities by those predicted from our
two-temperature modified blackbody fitting. Using the predicted the 500\,$\mu$m flux density increases the scatter in color space for some galaxies, and increases the $S_{350}/S_{500}$ ratios, illustrating the importance of
including submillimeter data when modeling the far-IR dust emission. \label{fig:color}}
\end{figure*}

\subsection{500\,$\mu$m excess when $\beta$ varies}
We calculate the excess emission at 500\,$\mu$m as
\begin{equation}
{\rm excess} = \frac{S_\nu (500\,\mu{\rm m}) - F_\nu (500\,\mu{\rm m})}{S_\nu (500\,\mu{\rm m})}
\label{eq:ex}
\end{equation}
where $F_\nu (500\,\mu{\rm m})$ is the predicted flux density of the two-temperature modified blackbody and $S_\nu$ is the measured flux density. We plot the excess emission as a function of metallicity in Figure \ref{fig:free_beta}. 
We correct the radius of each photometric region for the inclination of the galaxy, and then convert the radius to metallicity using the gradients listed in \citet{moustakas2010} which were calculated according to the 
\citet{pilyugin2005} relationship (see Table \ref{tbl:sample}). For M~101, which is not 
included in 
\citet{moustakas2010}, we use a metallicity gradient calculated directly from electron temperature measurements in H{\sc ii} regions \citep{kennicutt2003,bresolin2004}, 
which creates a slight offset between M 101 and the rest of our sample in metallicity.

Figure \ref{fig:free_beta} shows that there is no systematic dependence of the excess emission on metallicity for the sample as a whole, nor is there any trend within the individual galaxies that have a metallicity gradient. None of the galaxies display an excess larger than 25\%, and for many galaxies, the modeling actually overpredicts the 500\,$\mu$m emission by this amount, creating a 500\,$\mu$m deficit.
Furthermore, the spread in 
excess is largely accounted for by the uncertainty attached to each data point (the typical uncertainty for each region in the extended galaxies is shown in the bottom right of Figure \ref{fig:free_beta}). For some of the dwarf/irregulars, the uncertainties are rather large (particularly NGC 1377, NGC 2915, NGC 3773, NGC 5408, and IC 2574). The uncertainties are correlated with the SNR ratios of the SPIRE data since noisier photometry exacerbates the degeneracy between temperature and $\beta$ \citep{juvela2012}. The dwarf/irregulars with the lowest SNRs have the largest uncertainty on the derived $\beta$.

We find that no extended galaxy has an excess emission greater than 10\%, in agreement with the results of \citet{gordon2010} for a resolved study of the LMC using a modified blackbody model, and with \citet{aniano2013}, in which the authors create dust maps for the full sample of KINGFISH galaxies using the DL07 models. We show the distribution of excess emission
in the right panel of Figure \ref{fig:free_beta} for all regions in the extended galaxies, as well as the galaxies that were fit globally.
We find that the resolved elements of the extended galaxies do not preferentially show an excess or a deficit. The mean excess of our sample is $\sim0.02$. This confirms that the majority of regions display a low excess or deficit ($<10\%$). Given the large uncertainties (the dispersion is $\sim 0.1$), this excess is thus marginal.
In practice, we do not find excess 500\,$\mu$m emission greater than the scatter in the data.

\begin{figure*}
\centering
\plotone{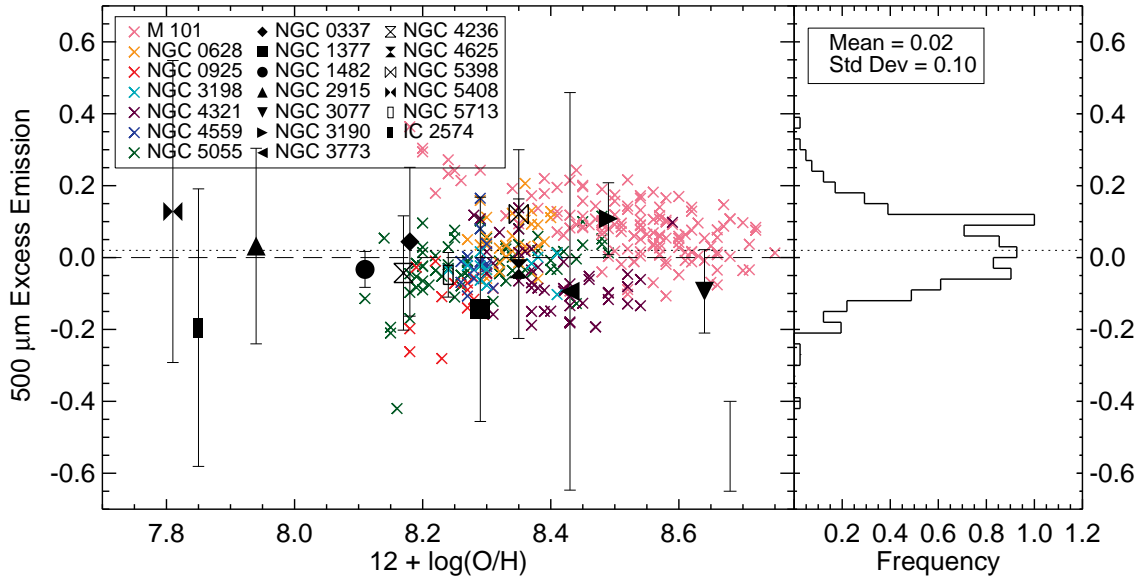}
\caption{Excess emission as a function of metallicity. We plot the excess for each individual region within an extended galaxy (colored crosses). The typical uncertainty is shown in the lower right corner. The dwarf/irregulars and 
compact galaxies are plotted as the black symbols. The galaxies with large uncertainties have low SPIRE SNRs. There is no trend in excess emission with decreasing metallicity for the extended galaxies or dwarf/irregular 
galaxies. On the right, we plot the distribution of the 500\,$\mu$m excess for all regions in the extended galaxies, as well as the galaxies that were fit globally; we have scaled the distribution to have a peak value of 1. 
The mean excess is $\sim2$\%. We indicate the mean with the
dotted line. The spike in excess above 2\% is largely due to NGC 5457 (M 101).\label{fig:free_beta}}
\end{figure*}

\subsection{500\,$\mu$m excess dependence on $\beta$}
The derived emissivity contains information about the slope of the Rayleigh-Jeans tail, and it is possible that $\beta$ is correlated with the measured excess emission \citep{galametz2012}
We test this hypothesis for our sample in
Figure \ref{fig:ex_beta}. For simplicity, we average together all of the derived excesses and $\beta$ values for each galaxy as we do not see any trends for individual galaxies.
There is no apparent correlation between the excess emission and the value of $\beta$ when $\beta$ is allowed to be a free parameter. However,
since $\beta$ and $T_c$ exhibit a degeneracy which is possibly enhanced by the particular fitting method used \citep[e.g.,][]{juvela2013}, a common technique for blackbody fitting is to hold $\beta$ fixed 
\citep[e.g.,][]{yang2007,xilouris2012}. It is possible that holding $\beta$ fixed will cause us
to measure an excess emission, since the slope of the Rayleigh-Jeans tail changes from region to region within each galaxy (see Figure \ref{fig:color}). We test this hypothesis by refitting all of our photometry holding 
$\beta$ fixed to the values of 1, 1.5, and 2.

\begin{figure}
\centering
\includegraphics[scale=0.47]{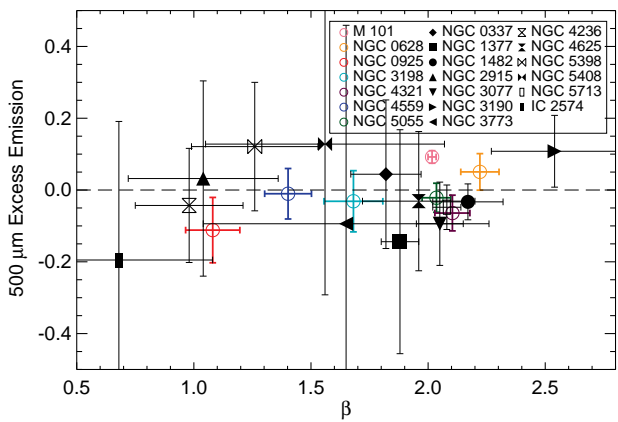}
\caption{The excess emission as a function of the derived emissivity. No obvious trend is evident, so allowing $\beta$ to vary does not produce any dependence between the derived $\beta$ and measured excess.
The large uncertainties on the emissivities are caused by an increased degeneracy between temperature and $\beta$ 
with low SNRs. \label{fig:ex_beta}}
\end{figure}

\begin{figure}
\centering
\includegraphics[scale=0.47]{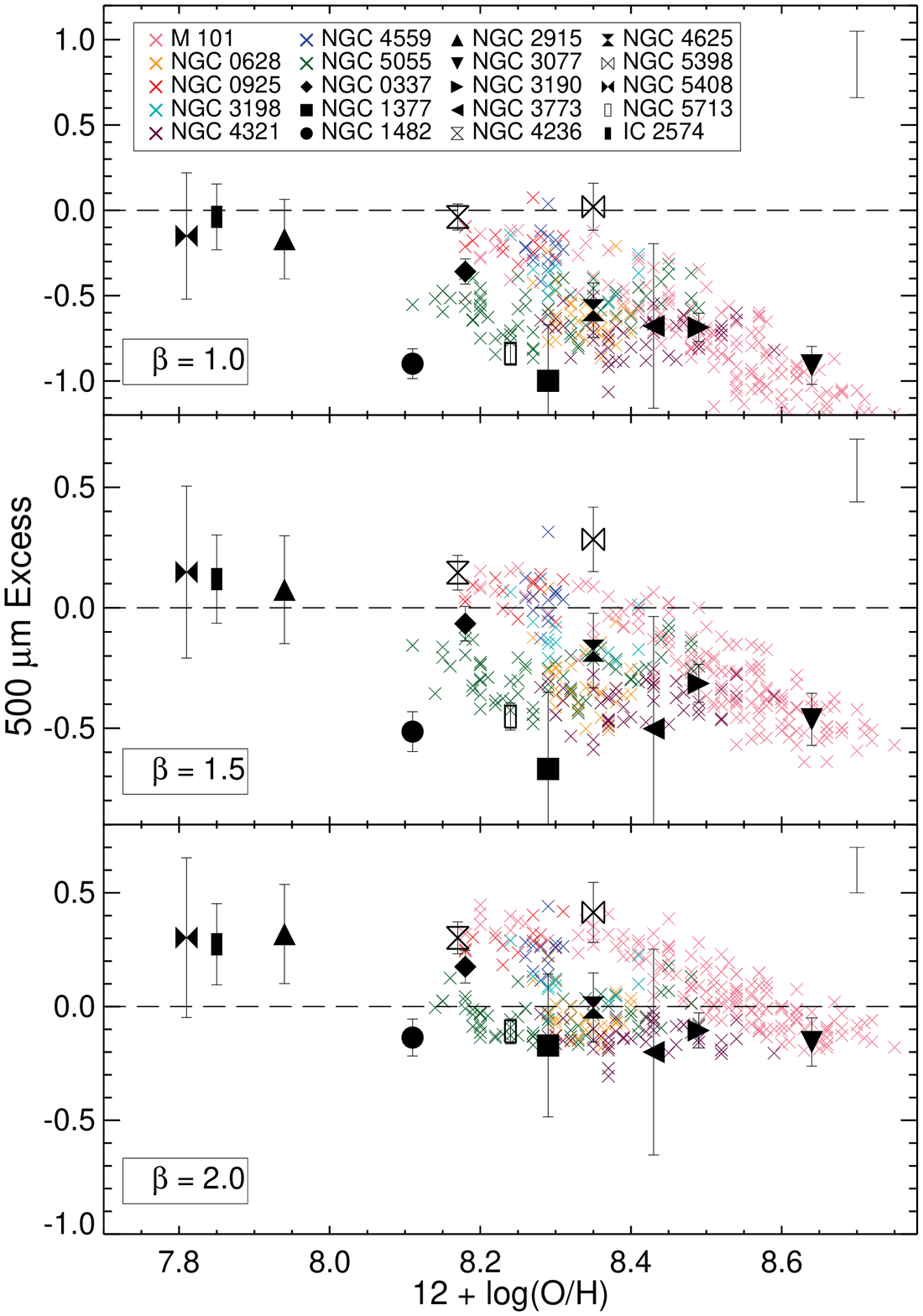}
\caption{We explore the effect of holding the emissivity constant on the measured excess emission. We set $\beta = 1$ (top panel), $\beta = 1.5$ (middle panel), and $\beta = 2$ (bottom panel). The typical uncertainty
for the regions in the extended galaxies is shown in the upper right corner of each panel.
When $\beta = 2$, we have the smallest uncertainties and best $\chi^2$ fits, whereas $\beta=1$ results in very poor $\chi^2$ values. \label{fig:fix_beta}}
\end{figure}

The 500\,$\mu$m excesses for each value of $\beta$ are plotted in Figure \ref{fig:fix_beta}. An emissivity of one generally produces very bad fits to the SEDs with high $\chi^2$ values for several galaxies, reflected in the error bars.
The slope of the Rayleigh-Jeans tail is too shallow and overestimates the 500$\,\mu$m emission, as can be seen in the top panel of Figure \ref{fig:fix_beta}. Setting 
$\beta=2$ provides the best fits to the SEDs when fixing the emissivity as can be seen by the small uncertainties in the bottom panel of Figure \ref{fig:fix_beta}. The excess emission is now $\sim$30\% for galaxies
which had only a 10\% excess when $\beta$ was allowed to vary (Figure \ref{fig:free_beta}). The result is consistent with \citet{remy2013}, who also calculate the excess 500\,$\mu$m emission keeping $\beta$ fixed at a value of 2. Holding $\beta$ fixed does increase the amount of excess emission measured. It also increases the so-called emission ``deficit",
meaning we now overpredict the 500\,$\mu$m emission by $\sim20-30\%$; the excess emission for each galaxy is listed in Table \ref{tbl:fit}. We stress that the amount of excess emission is affected by the particular fitting method used. In our study, where we achieve
good reduced $\chi^2$ fits when holding $\beta$ fixed to a value of two, and when letting $\beta$ be a free parameter, we find excess values differing by as much as 20\% depending on the fitting method.

There is a linear correlation between the excess emission and the metallicity. This is most strongly exhibited by M 101, which has a Pearsons correlation coefficient of $\rho=-0.86$ for all three values of $\beta$. NGC 4321
also has a strong trend between excess emission and metallicity. With the exception of NGC 5055, the other extended galaxies have fewer points, so it is difficult to see a trend.
The dwarf galaxies also display a high correlation with $\rho$ in the range $0.54-0.61$ depending on the value of $\beta$. On the other hand, NGC 4321 shows no correlation between metallicity and excess for any
value of $\beta$. When all of the galaxies are considered together, there is a strong anti-correlation ($\rho=-0.70$) for $\beta=1$ and only a low anti-correlation ($\rho = -0.30$) for $\beta=2$.

 IC 2574, NGC 2915, and NGC 5408 have a marked excess when $\beta=2$. However, for IC 2574 and NGC 2915, there is no substantial excess when $\beta = 1, 1.5$ or when $\beta$ is allowed to vary. From this, we
conclude that there is a flattening of the SED in these two galaxies that a steeper value of $\beta$ is unable to account for. Because $\beta$ is an effective emissivity, and contains information about the intrinsic properties of 
the dust grains as well as the heating of the dust within a resolution element, we cannot conclusively determine whether this flattening is due to changing properties of the dust grains in low metallicity galaxies.

For most galaxies, including the individual regions in the extended galaxies, the use of $\beta=2$ as opposed to $\beta=1$ produces much lower reduced $\chi^2$ values during the Monte Carlo fitting. We plot the 
distribution of reduced $\chi^2$ values over 1000 Monte 
Carlo fits for NGC 1377 in the top panel of Figure \ref{fig:chisq}. The mean
reduced $\chi^2$ is 2.3 when $\beta = 2$ but is 10.6 when $\beta = 1$. We also show the distribution of reduced $\chi^2$ values for NGC 2915, for which the choice of $\beta$ makes little difference in the goodness
of the fits. When $\beta=2$, the mean reduced $\chi^2 = 1.1$, and when $\beta = 1$, the mean reduced $\chi^2=1.3$. This effect is also seen for NGC 3773, NGC 5408, and IC 2574, all of which exhibit no decrease in the uncertainty on the excess depending on the value of $\beta$. These are the four galaxies with the lowest SNRs at 250\,$\mu$m and 350\,$\mu$m, illustrating the difficulty of using modified blackbody fitting with low signal-to-noise photometry.

\begin{figure}
\centering
\includegraphics[scale=0.57]{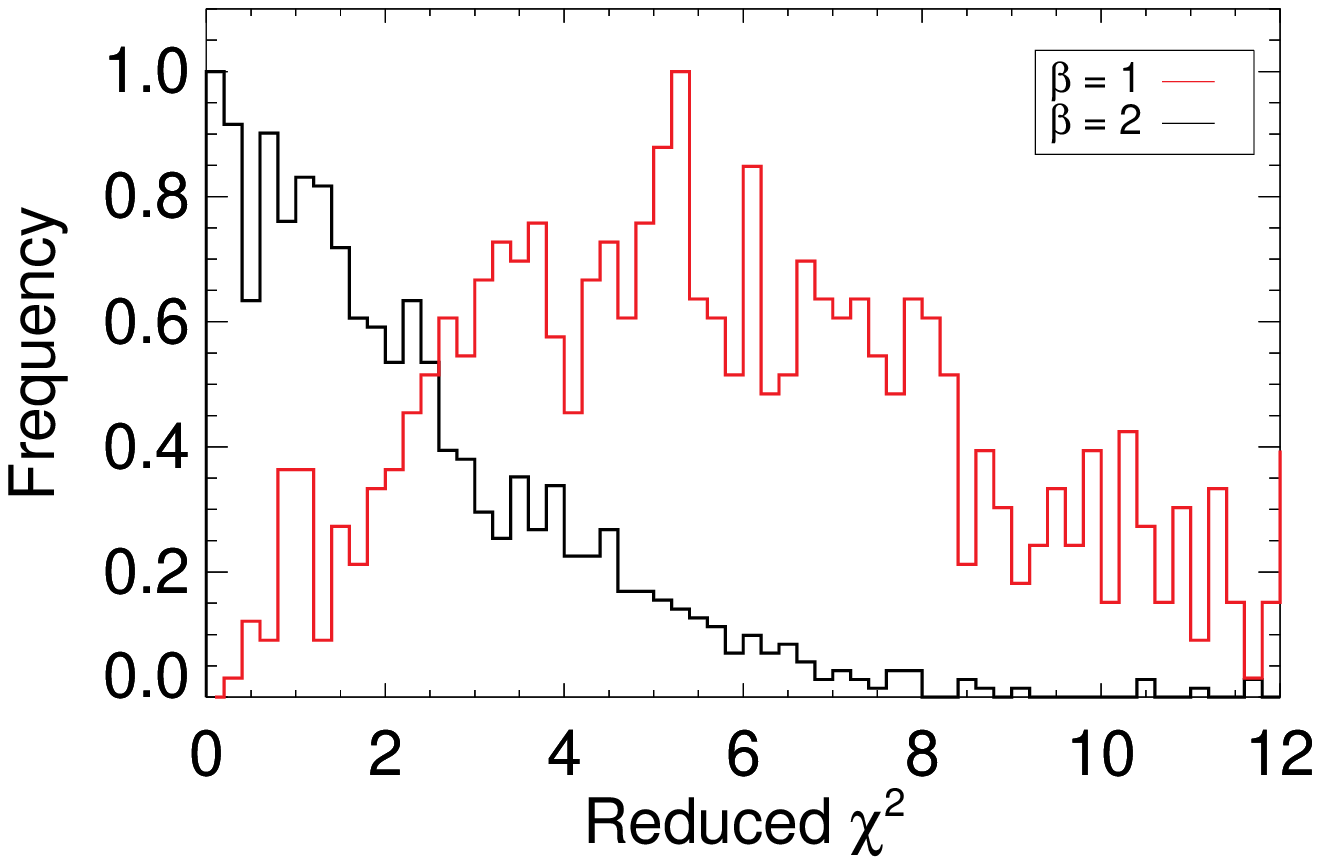}
\includegraphics[scale=0.57]{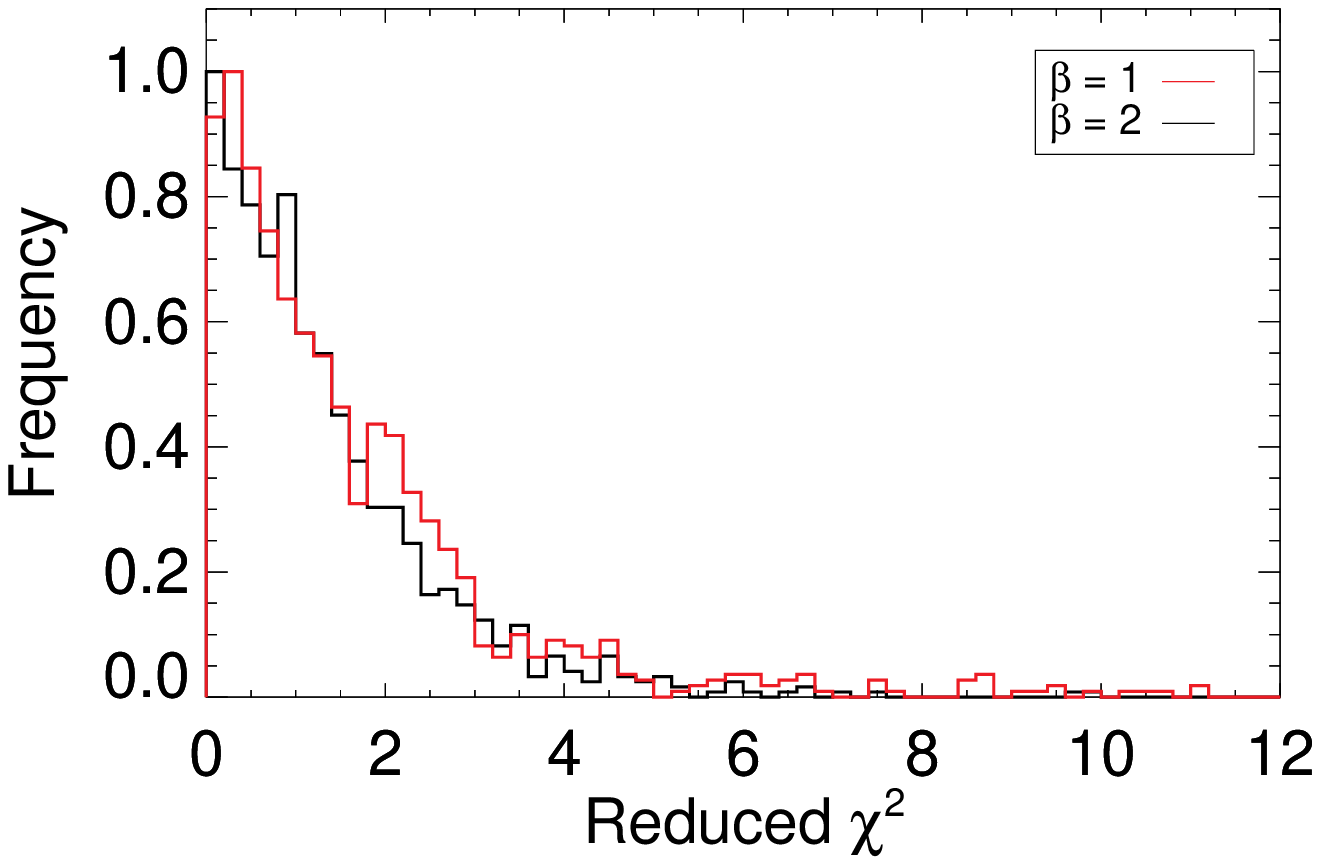}
\caption{The distribution of reduced $\chi^2$ values from each Monte Carlo repitiion when holding the emissivity fixed to a value of one (red) and two (black). For NGC 1377 (top), the choice of $\beta$ makes a large difference in the goodness
of the fits, but for NGC 2915 (bottom), it does not, though it does change the amount of excess emission calculated. \label{fig:chisq}}
\end{figure}

\subsection{Exploring the Submillimeter Dust Emission}
When $\beta$ is a free parameter, we do not see any systematic excess trend with metallicity, nor do we measure strong excess emission. However, when $\beta=2$, we see a weak dependence on metallicity, 
and we do detect some excess for individual galaxies (Table \ref{tbl:fit}).
One explanation for excess submillimeter emission is temperature mixing within a resolution element \citep[e.g.,][]{shetty2009}. The emissivity we measure is comprised both of the intrinsic emissivity of the dust and
the range of temperatures of the dust components in the resolution element. This range of temperatures can produce a shallower effective $\beta$ than would be measured in the ideal case of only one temperature 
component. We can test this explanation by comparing the excess emission with both the distance of our galaxies and the physical region size subtended by the photometry aperture (see Table \ref{tbl:sample}).
Both a larger region size and a larger distance will correspond to more dust emission within the resolution element, and hence more temperature mixing.
We find no trend between physical size and excess emission, demonstrating that the particular photometry aperture used does not affect the resulting excess.

In Figure \ref{fig:dist}, we plot the excess emission calculated when the emissivity was allowed to vary (see Figure \ref{fig:free_beta})
and when $\beta = 2$ (see Figure \ref{fig:fix_beta}) as a function of distance. There is a weak trend between the excess emission and the distance when $\beta=2$. When the emissivity is held fixed, galaxies that are within 10 Mpc have on average a higher excess
emission, whereas the 500\,$\mu$m emission has in general been overpredicted for the galaxies further than 10 Mpc. This does not appear to be an effect of galaxy type, since the nearby galaxies showing excess are 
both extended and dwarfs. The trend seen when $\beta=2$ is the opposite of what is expected. Galaxies that are further should have more temperature mixing within the resolution element
producing a shallower far-IR slope, which would cause
an excess when compared to the emission predicted with $\beta=2$. With the exceptions of NGC 0925 and NGC 4559, this trend is driven by the dwarf galaxies. Since we see no correlation between excess emission and
physical region size, or excess emission and distance when $\beta$ is a free parameter, we conclude that the excess emission is not due to temperature variations within the resolution element.

\begin{figure}
\centering
\plotone{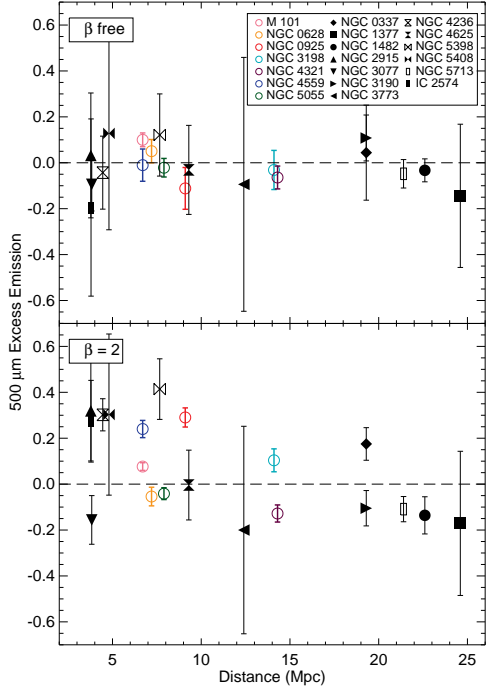}
\caption{The excess emission as a function of distance for the cases of a free emissivity (top see Fig. \ref{fig:free_beta}) and $\beta=2$ (bottom, see 
Fig. \ref{fig:fix_beta}). When the emissivity is held fixed, there is a trend with distance with the closer galaxies having more excess, and the farthest galaxies having on average a negative excess. \label{fig:dist}}
\end{figure}

NGC 2915, NGC 4236, NGC 5398, NGC 5408, and IC 2574 have
the highest excesses when $\beta = 2$. Multiple studies advocate using a third temperature component (T$\sim10$\,K) instead of a lower emissivity to account for the submillimeter excess
\citep{galametz2011,galliano2003,galliano2005,marleau2006}. We attempted to fit these four galaxies with a three temperature modified blackbody. We have only six photometric data points in each galaxy to
fit three modified blackbodies to, so we hold several parameters constant to avoid degeneracies. We set $T_w=60$\,K and $\beta_w=2$, as described in Section\ref{sec:fit}. We hold the emissivities of the cold
components fixed to a commonly applied value of 1.8, though we note that using $\beta=2$ caused our derived temperatures to change by only $\pm1\,$K. We hold the scaling of the warm dust component fixed to what we derived when fitting only two modified blackbodies. Finally, we restrict the temperature
range of one of the cold components to be $T\in[20,50]\,$K and the other to be $T\in[0,20]$\,K. We find that if we do not restrict the temperature ranges, then both cold components will have the same derived 
temperature.
We are able to achieve good fits (reduced $\chi^2 <1$) for NGC 5408 and NGC 5398 but not for NGC 2915 or IC 2574. For NGC 5408, our three dust temperatures are 60\,K, $29\pm3$\,K, and $13\pm7$\,K; for NGC 
5398,
they are 60\,K, $22\pm1$\,K, and $13\pm5$\,K. Also, it is important to note that not only are the uncertainties rather large on the coldest dust temperature, but the distribution of the derived temperatures for
the Monte Carlo simulations is not gaussian. There is only marginal excess ($<10\%$) when using three temperatures for these two galaxies, but given the large uncertainties, non-gaussian distribution, and the necessity of imposing
many restrictions on the fitting routine, we do not consider this method an improvement over using only two temperatures but allowing $\beta$ to vary.

\subsection{Comparison with Dale et al. (2012)}
D12 calculate the excess emission at 500 \,$\mu$m for all galaxies in the KINGFISH sample. They use a different approach and fit the physically motivated DL07 models and find that a dozen 
galaxies, including eight dwarf/irregular/Magellanic have an excess $>0.6$ at 500\,$\mu$m. We do not fit three of the dwarf/irregulars (Ho I, Ho II, and M81dwB) because we do not have 3$\sigma$ detections at all far-IR wavelengths. For the remaining dwarf/irregulars with substantial reported excess (NGC 2915, NGC 4236, NGC 5398, NGC 5408, and IC 2574)
we find at most excesses of $30$\% when $\beta$ is fixed at two. Another key difference in the two fitting approaches is that we are only concerned with fitting photometry spanning the range $24-500\,\mu$m, while
D12 fit all IR photometry from 3.6\,$\mu$m and long ward. Specifically, D12 had to balance emission from the stellar continuum, PAH features, and thermal dust emission, meaning that
they were unable to tailor their fits to the far-IR/submillimeter portion of the SED.

The discrepancy between the results of the present study and those of D12 can also be attributed to a difference in the photometry (Section \ref{sec:reduction}). 
The KINGFISH 500\,$\mu$m images used in D12
were recently recalibrated downward by of order 10\%. In addition, the PACS images were processed with different versions of the {\it Scanamorphos} package \citep{roussel2011}. In this work, all images were convolved to 
the 500\,$\mu$m resolution before measuring the photometry, whereas in D12, photometry was calculated using the original images. As noted in Section \ref{sec:background}, our background subtraction methods are different. It is also important to bear in mind that our photometry aperture sizes are not identical to D12. We use circular apertures whose diameters are listed in Table \ref{tbl:sample} while D12 uses elliptical apertures of varying sizes. Figure \ref{fig:dale} shows 
our photometry and the photometry for D12 overplotted. The main discrepancies are between the PACS data and the 500\,$\mu$m. D12 calculate the excess slightly differently than this work; namely, they divide the excess 
by the model flux density, whereas we use the measured flux density (Eq. \ref{eq:ex}).

We fit the D12 photometry for NGC 2915, NGC 4236, NGC 5398, NGC 5408, and IC 2574 with our two temperature modified blackbody model. We hold the emissivity fixed to a 
value of two, and we calculate the fractional excess according to the prescription in D12. For four of the galaxies, we find fractional excesses between 55\% and 75\% (IC 2574 has an excess of 44\%), leading us to conclude that discrepancy between our work and D12 is primarily due to the difference in the flux density calibration and convolution of the images, and the photometric apertures, and not a result of using different modeling techniques.

\begin{deluxetable*}{l | cccc|cc|cc|cc}
\tablecolumns{11}
\tablecaption{Derived Properties from SED fitting \label{tbl:fit}}
\tablehead{\colhead{} & \multicolumn{4}{c|}{$\beta$ Free} & \multicolumn{2}{c|}{$\beta=2$} & \multicolumn{2}{c|}{$\beta=1.5$} & \multicolumn{2}{c|}{$\beta=1$} \\
\hline
\multicolumn{1}{c|}{ID} & \colhead{(\%)\tablenotemark{a}} & \colhead{$\beta$} & \colhead{$T_c$ (K)} & \multicolumn{1}{c|}{Excess (\%)} & \colhead{$T_c$ (K)} & \multicolumn{1}{c|}{Excess (\%)}
& \colhead{$T_c$ (K)} & \multicolumn{1}{c|}{Excess (\%)} & \colhead{$T_c$ (K)} & \colhead{Excess (\%)}}
\startdata
NGC 0337 & 14 & 1.82 $\pm$ 0.15 & 23.8 $\pm$ 2.7 & \, \ 4 $\pm$ 21 & 23.7 $\pm$ 0.17 & 18 $\pm$ 7 \, &  26.9 $\pm$ 0.20 & -6 $\pm$ 7 & 31.3 $\pm$ 0.25 & -36 $\pm$ 7 \, \\
NGC 0628 & 11 & 2.17 $\pm$ 0.28  & 18.9 $\pm$ 1.6 & \ 5 $\pm$ 7 & 20.2 $\pm$ 0.83 & -5 $\pm$ 9 \ & 22.8 $\pm$ 1.0 \, & -31 $\pm$ 13 & 26.2 $\pm$ 1.2 \, & -62 $\pm$ 15 \\
NGC 0925 & 9   &  0.929 $\pm$ 0.41 \, & 24.4 $\pm$ 2.4 & -11 $\pm$ 8 \, & 19.5 $\pm$ 1.2 \, &  29 $\pm$ 6 \, & 21.9 $\pm$ 1.4 \, & \ 6 $\pm$ 6 & 25.0 $\pm$ 1.7 \, & -18 $\pm$ 9 \, \\
NGC 1377 & 16 & \, 1.89 $\pm$ 0.078 & 31.1 $\pm$ 2.8 & -14 $\pm$ 31 & 30.4 $\pm$ 0.73 & -17 $\pm$ 31 \ & 35.4 $\pm$ 0.89 & -67 $\pm$ 32 & 42.9 $\pm$ 1.2 \, & -100 $\pm$ 33 \, \\
NGC 1482 & 21 & 2.17 $\pm$ 0.15 & 24.4 $\pm$ 5.2 &  -3 $\pm$ 5    & 25.5 $\pm$ 0.21 &  -14 $\pm$ 8 \, \ & 29.5 $\pm$ 0.24 & -51 $\pm$ 8 \, & 35.2 $\pm$ 0.30 & -90 $\pm$ 9 \, \\

NGC 2915 & 7   & 1.04 $\pm$ 0.32 & 30.1 $\pm$ 3.2 & \, \ 3 $\pm$ 27   & 22.2 $\pm$ 0.79 &  32 $\pm$ 22  & 25.3 $\pm$ 0.90 & \, \ 8 $\pm$ 22   & 29.4 $\pm$ 1.1 \, & -17 $\pm$ 23 \\
NGC 3077 & 11 & 2.05 $\pm$ 0.10 & \, 26.8 $\pm$ 0.73 & \, -9 $\pm$ 12  & 27.1 $\pm$ 0.21 & -16 $\pm$ 11 \ & 31.2 $\pm$ 0.26 & -46 $\pm$ 11 & 37.1 $\pm$ 0.33 & -91 $\pm$ 11 \\
NGC 3190 & 17 & 2.54 $\pm$ 0.27 & 17.6 $\pm$ 2.9 & \ 11 $\pm$ 10 & 20.0 $\pm$ 0.19 & -11 $\pm$ 8 \, \ & 22.8 $\pm$ 0.22 &  -31 $\pm$ 8 \, &  26.7 $\pm$ 0.29 & -69 $\pm$ 8 \, \\
NGC 3198 & 14 & 1.41 $\pm$ 0.48 & 22.0 $\pm$ 3.3 &   -3 $\pm$ 4    & 19.3 $\pm$ 0.85 & 10 $\pm$ 9 \, & 21.7 $\pm$ 1.0 \, &  -12 $\pm$ 9 \, & 24.9 $\pm$ 1.3 \, & -39 $\pm$ 12  \\
IC 2574      & 31 & 0.683 $\pm$ 0.40 \, & 31.0 $\pm$ 5.8 & -20 $\pm$ 39 & 19.9 $\pm$ 0.88 &  27 $\pm$ 18 & 23.0 $\pm$ 1.0 \, & \ 12 $\pm$ 18 &  27.2 $\pm$ 1.4 \, & \, -4 $\pm$ 19 \\

NGC 3773 & 11 & 1.65 $\pm$ 0.61 & 25.1 $\pm$ 4.5 & \, -9 $\pm$ 55   & 22.9 $\pm$ 1.1 \, &  -20 $\pm$ 45 \ & 26.1 $\pm$ 1.3 \, & -50 $\pm$ 47 & 30.4 $\pm$ 1.6 \, & -68 $\pm$ 48 \\
NGC 4236 & 32 & 0.979 $\pm$ 0.23 \, & 23.1 $\pm$ 1.8 & \, -4 $\pm$ 16   & 17.6 $\pm$ 0.26 &  30 $\pm$ 7 \, & 19.9 $\pm$ 0.31 & 15 $\pm$ 7\, & 23.0 $\pm$ 0.40 & \, -4 $\pm$ 8 \, \\
NGC 4321 & 25 & 2.25 $\pm$ 0.56 & 18.2 $\pm$ 2.8 &  -6 $\pm$ 9   & 19.4 $\pm$ 1.4 \, & -13 $\pm$ 6 \, \ & 21.8 $\pm$ 1.6 \, &  -39 $\pm$ 8 \, & 25.0 $\pm$ 2.0 \, &  -73 $\pm$ 11 \\
NGC 4559 & 12 & 0.974 $\pm$ 0.51 \, & 25.6 $\pm$ 3.2 & -1 $\pm$ 7   & 20.3 $\pm$ 0.77 & 24 $\pm$ 8 \, & 22.8 $\pm$ 0.90 & \, \ 5 $\pm$ 10  & 26.1 $\pm$ 1.1 \, & -23 $\pm$ 11 \\
NGC 4625 & 8   & 1.96 $\pm$ 0.24 & 21.4 $\pm$ 1.4 & \, -3 $\pm$ 19   & 21.2 $\pm$ 0.32 & \, 0 $\pm$ 15 & 23.8 $\pm$ 0.36 & -18 $\pm$ 16 & 27.3 $\pm$ 0.44 & -59 $\pm$ 16 \\
NGC 5055 & 23 & 2.01 $\pm$ 0.35 & 18.5 $\pm$ 1.7 &  -2 $\pm$ 9    & 18.6 $\pm$ 1.4 \, & -4 $\pm$ 8 \ & 20.8 $\pm$ 1.6 \, &  -28 $\pm$ 10 &  23.7 $\pm$ 2.0 \, & -58 $\pm$ 13  \\
NGC 5398 & 17 & 1.26 $\pm$ 0.27 & 25.6 $\pm$ 2.1 & \ 12 $\pm$ 18  & 21.1 $\pm$ 0.47 & 41 $\pm$ 13 & 23.9 $\pm$ 0.56 & \ 28 $\pm$ 13 & 27.7 $\pm$ 0.66 & \, \ 2 $\pm$ 14 \\
NGC 5408 & 29 & 1.56 $\pm$ 0.51 & 30.5 $\pm$ 5.5 & \ 13 $\pm$ 42 & 26.5 $\pm$ 1.4 \,    & 30 $\pm$ 35 & 30.8 $\pm$ 1.7 \, & \ 15 $\pm$ 36 &  36.7 $\pm$ 2.2 \, & -15 $\pm$ 37 \\
M 101         & 11 & 2.02 $\pm$ 0.20 & \, 19.9 $\pm$ 0.36 & \ 9 $\pm$ 9   &  20.1 $\pm$ 0.58 & \, 7 $\pm$ 16    & 21.6 $\pm$ 0.78 & -27 $\pm$ 20  & 22.8 $\pm$ 0.73 & -78 $\pm$ 30 \\
NGC 5713 & 11 & \, 2.08 $\pm$ 0.063 & \, 24.0 $\pm$ 0.40 & -5 $\pm$ 6 & 24.4 $\pm$ 0.11 & -11 $\pm$ 6 \, \ & 28.0 $\pm$ 0.13 & -45 $\pm$ 6 \, &  32.8 $\pm$ 0.16 & -84 $\pm$ 6 \,
\enddata
\tablenotetext{}{For the 7 extended galaxies, all of the values listed in this table are averages and standard deviations of all of the photometric regions we modeled.}
\tablenotetext{a}{The percentage of the 100\,$\mu$m flux density due to the warm temperature modified blackbody (see Eq. \ref{eq:bb}).}
\end{deluxetable*}

\section{Conclusions}
We measure the emission at 500\,$\mu$m for a sample of star forming KINGFISH galaxies spanning a range of metallicities. We model the far-IR SED with a two temperature modified blackbody, and calculate
the excess emission as the measured 500\,$\mu$m emission relative to the predicted emission. For extended galaxies, we measure the excess in circular apertures with a diameter of 42'' from the center to the edge of the 
galaxy. We
supplement the sample of extended galaxies with a sample of compact and of dwarf/irregular galaxies, for which we model the far-IR emission on a global scale.

While we do see excesses of $\sim$\,10\% for a handful of galaxies when we allow $\beta$ to vary, we do not see any overall trend as a function of metallicity for the range $12 + \log($O/H$)=7.8-8.7$. Furthermore, this excess is completely
accounted for by the uncertainties in the fitting procedure. In addition, several of our galaxies exhibit a 500\,$\mu$m deficit of $\sim$\,10\%. We conclude that if any strong excess is present
at 500\,$\mu$m in lower metallicity galaxies, it must occur in galaxies with $12 + \log($O/H$) < 7.8$. Our findings are qualitatively consistent with the work of \citet{galametz2011}.
They find a submillimeter excess at 870\,$\mu$m in galaxies with $8 < 12 + \log($O/H)$<9$; however, when this excess is present, it is not universally seen at 500\,$\mu$m. The authors successfully account for the
excess with a very cold (T$\sim10$\,K) dust component. \citet{galametz2013} reports a submillimeter excess in some of the galaxies in this sample, but it is detected at 870\,$\mu$m. It is possible that such an
excess is present universally in our sample, but that longer wavelength observations are required to detect it. 

As for the flattening of the Rayleigh Jeans tail in the submillimeter regime, we find that $\beta=2$ most accurately reproduces the slope of SED around 500\,$\mu$m. In fact, using $\beta =1$, as has been suggested to
be appropriate at longer submillimeter wavelengths, consistently overestimates the emission at $500\,\mu$m. Therefore, any change in the emissivity does
not occur until $\lambda > 500\,\mu$m. An emissivity index of two
results in the lowest uncertainties in the SED fits and shows only a weak trend metallicity, which varies from galaxy to galaxy. Fitting with a standard value of $\beta=2$, which results in reduced $\chi^2 <1$, produces an excess of $\sim 30$\% in six galaxies, but over predicts emission by $\sim20\%$ in five galaxies. 
We conclude that though a small 500\,$\mu$m excess may be present in
individual galaxies, this is compensated by equal amounts of 500\,$\mu$m ``deficiency'' in comparable numbers of galaxies, and only preferentially affects the lower metallicity galaxies when modeling
with a fixed $\beta$.
\newline
\newline
Herschel is an ESA space observatory with science instruments
provided by European-led Principal Investigator consortia
and with important participation from NASA. IRAF, the
Image Reduction and Analysis Facility, has been developed by
the National Optical Astronomy Observatories and the Space
Telescope Science Institute.
This research has made use of the NASA/IPAC Extragalactic Database (NED) which
is operated by the Jet Propulsion Laboratory, California Institute of Technology, under
contract with the National Aeronautics and Space Administration.

\clearpage




\end{document}